\documentclass[a4paper,fleqn,usenatbib]{mnras}

\usepackage[T1]{fontenc}
\usepackage{ae,aecompl}

\usepackage{graphicx}	
\usepackage{amsmath}	
\usepackage{amssymb}	
\usepackage{soul}	
\usepackage{color,xcolor}

\title[physical properties of jets]{Kinetic powers of the relativistic jets in Mrk 421 and Mrk 501}

\author[Deng et al.]{
Xiao-Chun Deng,$^{1}$
Wen Hu,$^{1}$\thanks{E-mail: huwen.3000@jgsu.edu.cn}
Fang-Wu Lu$^{2}$\thanks{E-mail: sfweb@yxnu.edu.cn}
and Ben-Zhong, Dai$^{3}$\thanks{E-mail: bzhdai@ynu.edu.cn}
\\
$^{1}$Department of Physics, Jinggangshan University, Ji'an, 343009, Jiangxi, China\\
$^{2}$Department of Physics, Yuxi Normal University, Yuxi, 653100, China \\
$^{3}$Key Laboratory of Astroparticle Physics of Yunnan Province, Yunnan University, Kunming, 650091, China\\
}

\date{Accepted XXX. Received YYY; in original form ZZZ}

\pubyear{2020}

\begin{document}
\label{firstpage}
\pagerange{\pageref{firstpage}--\pageref{lastpage}}
\maketitle

\begin{abstract}
Using the standard one-zone synchrotron self-Compton (SSC) model and the Markov chain Monte Carlo (MCMC) technique,
we systematically analyze the quasi-simultaneous multi-wavelength (MWL) spectral energy distributions (SEDs)
of Mrk 421 and Mrk 501 during states of relatively low activity.
With this model in place, a semi-analytical method is developed to examine the uncertainty in jet power estimation caused by degeneracy of the radiative models.
The semi-analytical method, in combination with the MCMC technique, allows us to explore the jet properties over a wide range of the variability timescale.
Our results seem to support: (1) In both Mrk 421 and Mrk 501 the jets are powered by rapidly rotating black holes (BHs).
The BH spin in Mrk 501 could be lower than that in Mrk 421 or possibility they are equal under the assumption of Blandford-Znajek mechanism.
(2)The energy losses, which could be used to form the large-scale radio structure, are important for reconciling the differences of the kinetic power derived from the observations of the large-scale structure and the SED fitting results. Moreover, the jet energy losses in the propagation are more significant for Mrk 501 than for Mrk 421.
\end{abstract}




\section{Introduction}

Blazars are a relatively rare subclass of radio-loud active galactic nuclei (AGNs) with a relativistic jet pointing close to our line of sight \citep{Urry1995}.
According to the properties of their emission lines, blazars have been classified into BL Lacertae objects (BL Lacs) and flat-spectrum radio quasars (FSRQs).
Compared to FSRQs, BL Lacs show weak or absent optical emission lines \citep{Angel1980}.
The physical difference between these two subclasses may be connected to the different accretion model of the central
BH; that is, the accretion flow in FSRQs is in the regime of a standard disc accretion, while in BL Lacs it is in the advection-dominated accretion flow (ADAF)-like or
adiabatic inflow-outflow scenario (ADIOS)-like regimes \citep[see][]{GT2008,Ghisellini2009}.

The emission from blazars covers entire electromagnetic spectrum, and is dominated by non-thermal and variable radiation originating from the innermost part of jet
that is well beyond the current imaging capabilities of telescopes in any part of the electromagnetic spectrum.
The multi-wavelength (MWL) spectral energy distributions (SEDs) of blazars are usually composed of two bumps.
The peak of the first bump ranges between infra-red and X-rays, whereas the second one shows in the $\gamma$-rays \citep[e.g.,][]{Abdo2010}.
It is commonly accepted that the SEDs of BL Lacs are produced by synchrotron and synchrotron-self-Compton of a non-thermal population of electrons
\cite[SSC model; e.g.,][]{Konigl1981,Bloom1996,Mastichiadis1997}.
This scenario is favored by the tight X-ray and very high energy (VHE) $\gamma$-ray correlation and the very rapid $\gamma$-ray variability observed in BL Lacs \citep[e.g.,][]{Gliozzi2006,Albert2007,Fossati2008,Acciari2011,Furniss2015,Aleksi2015a,Aleksi2015b,Balokovi2016}.
For FSRQs, the SEDs need to invoke comptonization of the external photon field surrounding the jets \citep[e.g.,][]{Sikora1994,Blazejowski2000,Bottcher2002,Ghisellini2010,D2009}.

Radiative properties make blazars ideal laboratories to understand the physics of relativistic jet, including jet launching, energy transportation, energy dissipation and conversion processes, etc.
Using the MWL SEDs models, the jet powers as well as the magnetization and radiative efficiency can be inferred, which are fundamentally important in studying the jet physics \citep[e.g.,][]{Celotti1993,Celotti2008,Zhang2012,Zhang2013,Zhang2014,Chen2018,Fan2018}.
However, the modelling results are strongly dependent on the plasma composition of the jets, which is largely unknown.
Although a pure $e^\pm$ pairs composition in FSRQs is disfavored because the Compton rocket effect would stop the jet \citep{Sikora2000,GT2010,Ghisellini2012}, it cannot be excluded in BL Lacs \citep{Petropoulou2019}. 
With assumption of one proton per electron, the jet power of BL Lacs inferred from the fitting results of the SED is significantly higher than the kinetic power that derived from the observations of the large-scale structures, and is comparable or even exceeds the accretion power \citep[e.g.,][]{Ghisellini2014,Madejski2016}.
Recently, a mixed composition with tens of pairs per proton has been proposed to reconcile the discrepancy between the jet powers estimated by the two methods \citep{Kang2014,Sikora2016,Pjanka2017,Fan2018}.

In this paper, we investigate the radiative properties of BL Lacs in the framework of a SSC model. Different from the previous works in the literature, we focus our attention on the energy losses from pc to kpc scales during the growth of large-scale radio structures of BL Lacs, and then both $e^\pm$ pair and electron-proton compositions are discussed.
Meanwhile, the jet formation mechanism of BL Lacs is explored according to the comparisons of the jet power derived from SED fitting and the accretion power of the central BH. It should be noted that, to reduce the uncertainties both on the occasional flaring activities and on the radiative mechanisms, the investigations in this paper are focused on the average/typical properties of the objects.
The modelling of the SEDs of BL Lacs generally suffer from the inherent degeneracies of the radiative models, so a semi-analytical method is developed to clarify the influence of the uncertainty and the Markov chain Monte Carlo (MCMC) method is employed to obtain the best-fitting result and the associated uncertainty.
Specifically, we present a first application to the high-synchrotron-peaked BL Lacs (HBLs) Mrk 421 and Mrk 501.
The paper is structured as follows. A brief description of the method is presented in Section \ref{sec:intro}.
The results and discussion are shown in Section \ref{sec:result}.
The conclusions are given in Section \ref{sec:summa}.

Throughout the paper, the luminosity distance $d_L$ was computed with the online tool CosmoCalc\footnote{\url{http://www.astro.ucla.edu/~wright/CosmoCalc.html}} \citep{w06} using a standard flat cosmology with $\rm H_0 = 69.6\,$km/s/Mpc, $\rm \Omega_M = 0.286$ and $\Omega_\Lambda=0.714$.

\section{Method} \label{sec:intro}

In the standard one-zone leptonic jet model, the blazar emission zone is assumed to be a spherical blob of radii $R^\prime$ composed of a
uniform magnetic field ($B^\prime$) and a population of isotropic relativistic electrons of density $n_{\rm e}'$.
The blob propagates with a relativistic speed $\beta_\Gamma = (1-1/\Gamma^2 )^{1/2}$ (normalized by the speed of light $c$) outward along the jet, which is directed at an angle $\theta\sim1/\Gamma$ with respect to the line of sight.
Thus, the observed radiation are strongly boosted by a relativistic Doppler factor given by $\delta_{\rm D} = 1/[\Gamma(1-\beta_\Gamma \cos\theta)]$.
Moreover, the jet kinetic power $L_{\rm kin}$, magnetization $\sigma_{\rm m}$
\footnote{Note that the definition of $\sigma_m$ can be written in form of $U_B'/U_e'$ or $U_B'/(U_e'+U_p')$.
It could be equivalent to the definition of $\sigma \equiv B'^2/4\pi\rho c^2$, where both the magnetic field strength $B'$ and the rest-mass density $\rho$ are measured in the rest frame of the fluid \citep{Sironi2015}}, and radiative efficiency $\eta_{\rm r}$ can be expressed as
\begin{eqnarray}
  L_{\rm kin} &=& L_{\rm B} + L_{\rm e} + \eta_pL_{p,n_{\rm p}'=n_{\rm e}'},\\
  \sigma_{\rm m} &=& L_{\rm B}/(L_{\rm kin}-L_{\rm B}), \\
  \eta_{\rm r} &=& L_{\rm r}/(L_{\rm kin}+L_{\rm r}),
\end{eqnarray}
where $\eta_{\rm p}\equiv n_{\rm p}'/n_{\rm e}'$ denotes the physical number ratio of the protons $(n_{\rm p}')$ to electrons $(n_{\rm e}')$, %
$L_{\rm B}$, $L_{\rm e}$ and $L_{\rm r}$ are the jet powers carried by magnetic field, electrons and radiation field, respectively,
and $L_{\rm p,n_p'=n_e'}$ is the protonic power of a normal jet with pure $e^--p$ plasma, i.e. $\eta_{\rm p}=1$.
On the other hand, $\eta_{\rm p}=0$ corresponds to a jet with pure $e^\pm$ pairs.

With the assumption of one proton per emitting electron, the two-sided jet powers carried by each ingredient ($L_{\rm B}, L_{\rm e}, L_{\rm p}$ and $L_{\rm r}$)
can be given by \citep[e.g.][]{Celotti1993,Celotti2008}
\begin{equation}\label{jpower}
L_i=2\pi {R'}^2\Gamma^2\beta_\Gamma cU_i^\prime,
\end{equation}
where $U'_i(i= {\rm B,e,p,r})$ are the energy densities associated with the magnetic field $U_{\rm B}'=B'^2/8\pi$,
the radiating electrons $U_{\rm e}'=m_{\rm e}c^2\int\gamma'n_{\rm e}'(\gamma')d\gamma'$, the cold protons $U_{\rm p}'=m_{\rm p/e}U_{\rm e}'/\overline\gamma'$,
and the radiation field $U_{\rm r}'=L_{\rm r,obs}/4\pi{R'}^2c\delta_{\rm D}^4$ in the comoving frame, respectively.
Here, $m_{\rm p/e}$ is the ratio of the rest mass of proton to electron,
$L_{\rm r,obs}$ is the total observed non-thermal luminosity and $\overline\gamma'=\int\gamma'n_{\rm e}'(\gamma')d\gamma'/\int n_{\rm e}'(\gamma')d\gamma'$ represents the average Lorentz factor of relativistic electrons.

In this study, the energy distribution of the electrons in the blob is directly assumed to be a power-law function with an exponential cutoff (PLC) of index $\beta$,
and is thus described as
\begin{equation}\label{plc_eed}
n_{\rm e}^\prime(\gamma^\prime) = N_{\rm 0}^\prime\left(\frac{\gamma^\prime}{\gamma_{\rm c}^\prime}\right)^{-\alpha}\exp\left[-\left(\frac{\gamma'}{\gamma_{\rm c}'}\right)^{\beta}\right],~\gamma_{ l}^\prime\le\gamma^\prime < \gamma_{\rm u}^\prime
\end{equation}
where $N_0'$ is the normalization factor, $\gamma'_{l/\rm {c/u}}$ is the minimum/critical/maximum Lorentz factor of relativistic electrons, $\alpha$ denotes the spectral index at $\gamma'\ll\gamma_{\rm c}'$ \citep{Lefa2012}.
We notice that the cutoff index $\beta$ allows us to describe a quite broad range of distributions, even vary sharp, abrupt, step-function like cut-off.
In principle, this distribution could be a natural outcome of the process of the first-order Fermi acceleration \citep{Webb1984,Dempsey2007}.

\subsection{Theoretical SED of non-thermal photons}

For a given electron energy distribution (EED), the observed synchrotron spectrum of the blob in the $\nu-\nu F_\nu$ diagram can be given by
\begin{equation}
f_\epsilon^{\rm syn}=\frac{\delta_{\rm D}^4\sqrt{3}e^3B^\prime}{4\pi hd_{\rm L}^2}\chi(\tau_{\epsilon^\prime})\epsilon^\prime V_{\rm b}'\int_1^\infty d\gamma^\prime n_{\rm e}^\prime(\gamma^\prime) R_{\rm s}(\epsilon^\prime/\epsilon^\prime_{\rm c}),
\end{equation}
where $e$ is the fundamental charge, $h$ is Planck's constant, $V_{\rm b}^\prime=4\pi {R^\prime}^3/3$ is the intrinsic volume of the blob, $d_{\rm L}$ is the luminosity distance of the source at a redshift of $z$, and $\chi(\tau)=3u(\tau)/\tau$ is defined as the Synchrotron self-absorption (SSA) factor, and $u(\tau)=1/2+\exp(-\tau)/\tau-[1-\exp(-\tau)]/\tau^2$, with the opacity $\tau=2\kappa_{\epsilon^\prime} R'$ \citep{DM2009}.
In the equation, the function $R_{\rm s}(x)$ is the monochromatic emission power
averaged over a population of electrons with randomly distributed pitch angle \citep{crusius1986}, and an accurate approximation given by \cite{Finke2008} is adopted in the calculation.

The dimensionless SSA coefficient is given by
\begin{equation}
\kappa_{\epsilon^\prime}=\frac{-\sqrt{3}B^\prime\lambda_{\rm c}^3e^3}{8\pi hm_{\rm e}c^3{\epsilon^\prime}^2}\int_1^\infty d\gamma^\prime R_{\rm s}\left(\frac{\epsilon'}{\epsilon'_{\rm c}}\right)\Big[{\gamma'}^2\frac{\partial}{\partial\gamma^\prime}\Big(\frac{n_{\rm e}^\prime(\gamma^\prime)}{{\gamma^\prime}^2}\Big)\Big],
\end{equation}
where $\lambda_{\rm c}=h/m_{\rm e}c=2.43\times10^{-10}~\rm cm$ is the electron Compton wavelength,
and $\epsilon_{\rm c}^\prime={3eB^\prime h}{\gamma^\prime}^2/{4\pi m_{\rm e}^2c^3}$ is the dimensionless characteristic energy of synchrotron radiation.

In the $\nu-\nu F_\nu$ diagram, the observed SSC spectrum is given by \citep[e.g.,][]{Jones1968,Blumenthal1970,D2009}
\begin{equation}\label{forma:ssc}
f_{\epsilon_\gamma}^{ssc}=\frac{{3c\sigma_TV_b^\prime}\delta_D^4}{16\pi d_L^2}\epsilon_\gamma'^2\int_0^\infty{}d\epsilon' \frac{u_{syn}'(\epsilon')}{\epsilon'^2}\int_{\gamma_l'}^{\gamma_u'} {}d\gamma^\prime{}\frac{n_e'(\gamma')}{\gamma'^2}F_{c}(x,q),
\end{equation}
where $\sigma_{\rm T}$ is the Thomson cross section, the energy densities of synchrotron radiation can be calculated through
$u_{syn}'(\epsilon')=\frac{3 d_L^2f_\epsilon^{syn}}{R^{\prime2}c\epsilon^\prime\delta_D^4}$, and
\begin{eqnarray}
  F_{\rm c}(x,q) &=& \Big[2q\ln{q}+q+1-2q^2 + \frac{(xq)^2}{2(1+xq)}(1-q)\Big] \nonumber\\
   &\times& H(q;\frac{1}{4\gamma'^2},1),
\end{eqnarray}
where
\begin{equation}
 q=\frac{\epsilon_\gamma'/\gamma'}{x(1-\epsilon_\gamma'/\gamma')},~~~~x=4\epsilon'\gamma'.
\end{equation}

Using the relation $n_e^\prime(\gamma^\prime) =\frac{6\pi d_L^2 f^{syn}_{\epsilon_e,\delta} }{c\sigma_TV_b' U_B'\gamma^{\prime3}\delta_D^4}$ (see Appendix \ref{appenda}),
and transforming the integral over $\gamma'$ in Eq.\ref{forma:ssc} to an integral over synchrotron photon energy $\epsilon_e$ according to the relations
$\gamma'=\sqrt{\frac{1+z}{\delta_D}\frac{3\epsilon_e}{4\epsilon_{\rm B}'}}$ and $d\gamma'=\sqrt{\frac{1+z}{\delta_D}\frac{3}{4\epsilon_{\rm B}'}}\frac{d\epsilon_e}{2\epsilon_e^{1/2}}$,
give
\begin{equation}\label{ssc_spectra}
  f_{\epsilon_\gamma}^{\rm ssc}=\left(\frac{d_{\rm L}/R'}{1+z}\right)^2\mathcal{F}_{\epsilon_\gamma}^{\rm ssc},
\end{equation}
where the function $\mathcal{F}_{\epsilon_\gamma}^{\rm ssc}$ is described by
\begin{equation}\label{source_flux}
  \mathcal{F}_{\epsilon_\gamma}^{\rm ssc}=\left[\frac{24\pi\epsilon_\gamma^2}{cB_{\rm cr}^2\delta_{\rm D}^2}
   \int_{\rm 0}^\infty{}d\epsilon \frac{f_{\epsilon}^{\rm syn}}{\epsilon^3}\int_{\epsilon_l}^{\epsilon_{\rm u}} d\epsilon_{\rm e}\frac{f_{\epsilon_{\rm e},\delta}^{\rm syn}}{\epsilon_{\rm e}^3}F_{\rm c}(x,q)\right].
\end{equation}

Here, $\epsilon_{\rm B}'=B'/B_{\rm cr}$ is the ratio of $B'$ and the critical magnetic field $B_{\rm cr}\simeq4.41\times10^{13} ~{\rm G}$,
$f_{\epsilon_{\rm e},\delta}^{\rm syn}$ is the $\delta$-function approximation for the synchrotron radiation, and
 $\epsilon$ and $\epsilon'$ refer to the dimensionless photon energy in the observer and comoving frame, respectively.
The photon energies measured in the two frames are related through $\epsilon=\epsilon'\delta_D/(1+z)$.

\subsection{Fitting strategy and procedure}\label{strategy}

To efficiently reconstruct the observed synchrotron bump, we set the synchrotron peak frequency $\nu_{\rm pk}$ and peak flux $f_{\rm pk}^{\rm syn}$ as the free parameters instead of $N_{\rm 0}'$ and $\gamma_{\rm c}'$, and is thus expressed as
\begin{eqnarray}
  \gamma_{\rm c}' &=& \sqrt{\frac{\nu_{\rm pk}}{\nu_{\rm 0}B'\delta_{\rm D}}}\phi \label{gamb1}\\
  N_{\rm 0}' &=& \frac{ f_{\rm pk}^{\rm syn}}{f_{\rm 0}V_{\rm b}'B'^2\gamma_{\rm c}'^3\delta_{\rm D}^4}\psi, \label{n01}
\end{eqnarray}
where $\nu_0$ and $f_0$ are constant (see Appendix \ref{appenda}), and $\phi$ and $\psi$ in our considered EED can be given by
\begin{eqnarray}
  \phi &=& [(3-\alpha)/\beta]^{-1/\beta}, \\
  \psi &=& [(3-\alpha)/\beta]^{(\alpha-3)/\beta}\exp\left[(3-\alpha)/\beta\right],
\end{eqnarray}
respectively. Note that $\phi$ and $\psi$ hold for $\alpha<3$. Furthermore, we parameterize $\gamma_l'$ and $\gamma_{\rm u}'$ relative to $\gamma_{\rm c}'$ by introducing a dimensionless parameters,
$\eta_{l/{\rm u}}=\gamma_{l/{\rm u}}'/\gamma_{\rm c}'$.
Here, $\eta_l$ and $\eta_{\rm u}$ are set as input parameters instead of $\gamma_{l}'$ and $\gamma_{\rm u}'$.
These input parameters could be determined predominantly by the width and slope of the observed spectrum below and above the synchrotron peak.

The model adopted is thus characterized by 9 parameters: $\nu_{\rm pk},~f_{\rm pk}^{\rm syn},~\alpha,~\beta,~\eta_l,~\eta_{\rm u},~B',~\delta_{\rm D}$ and $R'$.
Since the model is not sensitive to $\eta_{\rm u}$, the parameter $\eta_u$ is fixed to be $10^2$ in the calculation.
According to the SSC model, the spectral indices for the two bumps are consistent, and the peak frequency and flux provided by the observed Compton bump
are not adequate to uniquely determine the global properties of the emission zone, i.e. $B',~\delta_{\rm D}$ and $R'$.
However, we notice that $R'$ chiefly controls the peak flux of the Compton bump, and is independent on it's peak frequency.
For a given value of $R'$ \footnote{Notice that a plausible value for $R'$ could be pre-selected based on the previous work or independent method.},
a solution can then be uniquely determined  by using a $\chi^2$ minimization method to fit emission model curves to the observational data.
Subsequently, the jet properties including the jet powers, magnetization and radiative efficiency can be obtained for a given value of $\eta_p$.

In the fitting strategy of the observed SEDs of BL Lacs, to more efficiently determine the best values and associated uncertainties of the remaining parameters,
a MCMC fitting technique based on Bayesian statistics is employed to explore the multi-dimensional parameters space systematically \citep[e.g.,][]{Yan2013,Yan2015,Qin2018}.
To further consider the uncertainties on the transport parameters, the confidence intervals of physical quantities of interest are also obtained by using the MCMC code.
The more details on the MCMC technique can be found in the literature \citep[e.g.][]{Lewis2002,Yuan2011,Liu2012}.

Finally, to clarify the influence of $R'$ on the jet properties of BL Lacs, we develop a semi-analytical approach
to find a family of solutions that could be equally applicable to describe the observed SED at hand.
This approach is discussed in detail in the Appendix. \ref{para_space_study}.

\begin{table*}
\def\arraystretch{1.5}
\caption{The best-fitting values of the input parameters for Mrk 421 and Mrk 501.}
\label{tab:source1}
\begin{tabular}{lccccccccc}
\hline
Model & $\nu_{\rm pk}$	  & $f_{\rm syn}^{\rm pk}$	& $\eta_l$ & $\alpha$	&  $\beta$  & $B'$  & $\delta_{\rm D}$ &$\chi_{\rm redu}^2$  \\
     & $\rm 10^{17} ~Hz$ & $\rm 10^{-11} ~erg/cm^2/s$	& $10^{-3}$	&  &  & $10^{-2}$  &  &  \\
\hline
A1  & $1.16_{-0.21}^{+0.28}$ & $34.87_{-1.85}^{+1.41}$	& $3.72_{-0.90}^{+2.08}$  &$2.22_{-0.18}^{+0.12}$ &$1.04_{-0.25}^{+0.28}$& $4.75_{-0.85}^{+1.60}$ &$51.68_{-8.30}^{+6.00}$ &1.8\\
A2  & $1.27_{-0.23}^{+0.41}$ & $34.22_{-1.73}^{+1.63}$	& $3.22_{-0.69}^{+1.24}$  &$2.30_{-0.14}^{+0.14}$ &$1.16_{-0.25}^{+0.44}$& $2.27_{-0.52}^{+0.67}$ &$21.03_{-3.04}^{+3.73}$ &2.1\\
A3  & $1.51_{-0.33}^{+0.47}$ & $33.53_{-1.58}^{+1.78}$	& $2.81_{-0.44}^{+0.74}$  &$2.40_{-0.13}^{+0.11}$ &$1.43_{-0.35}^{+0.53}$& $1.04_{-0.22}^{+0.26}$ &$8.70_{-1.14}^{+1.52}$ &2.5 \\
\hline
A1  & $1.88_{-0.21}^{+0.88}$ & $5.51_{-0.22}^{+0.26}$	& $8.80_{-7.94}^{+1.19}$  &$2.15_{-0.08}^{+0.36}$ &$0.38_{-0.02}^{+0.35}$& $1.58_{-0.97}^{+1.22}$ &$51.58_{-13.45}^{+33.11}$ &0.9\\ 
A2	& $2.29_{-0.59}^{+1.13}$ & $5.51_{-0.25}^{+0.51}$	& $1.45_{-0.64}^{+9.18}$  &$2.41_{-0.34}^{+0.50}$ &$0.59_{-0.23}^{+0.42}$& $0.61_{-0.23}^{+0.80}$ &$24.59_{-8.07}^{+13.69}$ &0.9\\
A3	& $2.38_{-0.70}^{+0.85}$ & $5.51_{-0.29}^{+0.26}$	& $1.39_{-1.12}^{+8.60}$  &$2.42_{-0.35}^{+0.19}$ &$0.60_{-0.23}^{+0.32}$& $0.33_{-0.16}^{+0.28}$ &$9.37_{-2.83}^{+3.77}$ &0.9\\
\hline
\end{tabular}
\begin{flushleft}
Note: The upper and lower parts are for Mrk 421 and Mrk 501, respectively. Model A1, A2 and A3 correspond to $\rm R^\prime=10^{16}, 10^{17}$ and $10^{18}$ cm.  \\
\end{flushleft}
\end{table*}

\begin{figure*}
  \centering
  \includegraphics[width=0.5\textwidth]{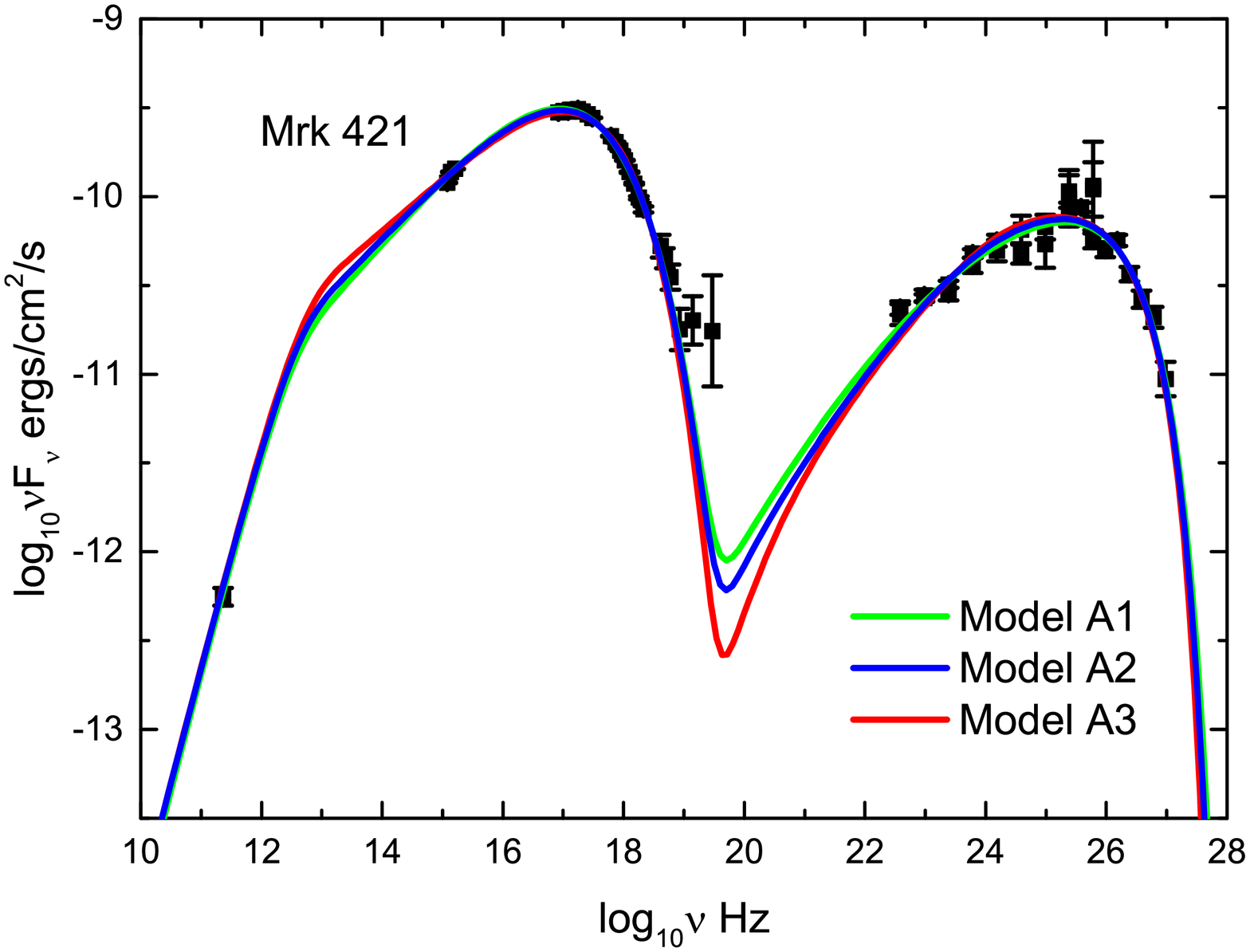}\includegraphics[width=0.5\textwidth]{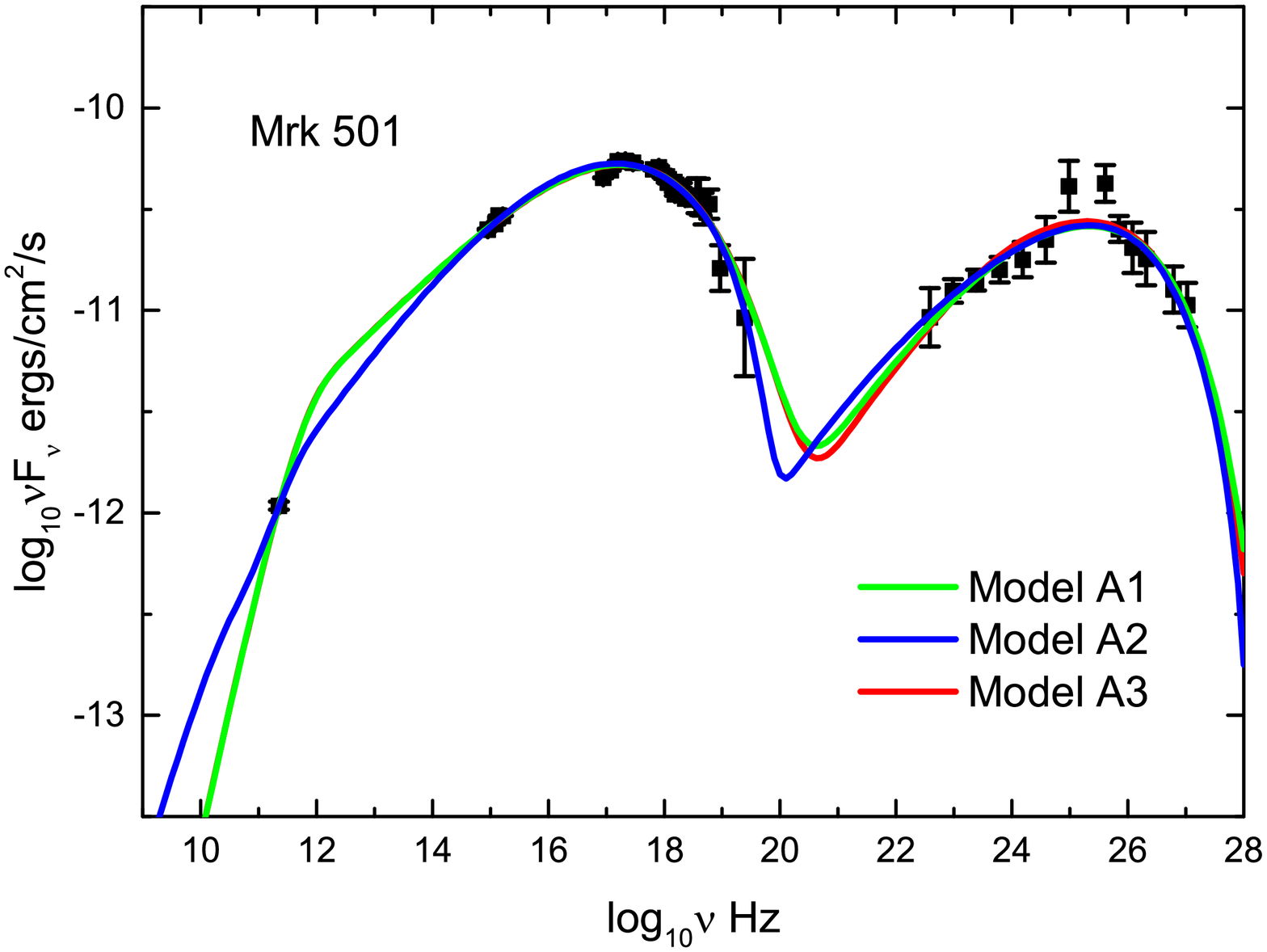}
  \caption{Modeling the non-thermal emission from the radio to TeV-ray bands for Mrk 421(left) and Mrk 501(right).
 }\label{sed_fitting}
\end{figure*}

\section{Results and discussion}\label{sec:result}

In this section, we now apply the model to the two famous HBLs: Mrk 421 and Mrk 501.
The quasi-simultaneous MWL SEDs of these two objects with nicely sampling from radio up to TeV $\gamma$-rays are collected from \cite{Abdo2011b} and \cite{Abdo2011a}.
In the radio band, we only take the \textmd{SMA} data at 230 GHz into account. Meanwhile, the $\gamma$-rays data observed by \emph{Fermi}-LAT and \textmd{MAGIC}
and the optical-UV to X-ray data observed by \emph{Swift}/UVOT/RXT/BAT are adopted in this paper.
For Mrk 501, the X-rays data from \emph{RXTE}/PCA are also taken into account.
In the fitting, a relative systematic uncertainty of 5\% was added in quadrature to the statistical error of the
Radio-Optical-UV-X-rays data, as usually adopted in the literature \citep[e.g.,][]{Poole2008,Abdo2011a,Wu2018}.

Using the fitting strategy and procedure shown in section \ref{strategy}, the comparisons of the best-fitting SEDs with the observational data set for Mrk 421 and Mrk 501 are plotted in Figure \ref{sed_fitting}.
To keep the range of allowed model parameter values as broad as possible and to reduce the degeneracy of the radiative model,
we respectively perform the SED fitting of the radius of the blob with three given values of $R'= 10^{16}, 10^{17}$ and $10^{18}$ cm,
which are referred as Model A1, A2, A3, respectively.
The best-fitting values of the parameters with 68\% errors and the reduced $\chi^2_r$-values are listed in Table \ref{tab:source1}.
The modelling results indicate that the unprecedented, complete SEDs for both Mrk 421 and Mrk 501 can be successfully fitted,
but it is difficult to distinguish among the three models (A1, A2, and A3) based on their $\chi^2_r$-values.

Note that the high $\delta_D-$values deduced in Model A1 may not be consistent with the fact that the two HBLs show almost no superluminal motion in the Very Long Baseline Array (VLBA) scale \citep[e.g.,][]{Piner2004,Giroletti2006}. However, the high $\delta_D$-values are required to account for the observed rapid $\gamma-$ray variability presented in previous studies \citep[e.g.,][]{Albert2007,Paliya2015}.
This discrepancy indicates that the jet may either undergo severe deceleration \citep{Georganopoulos2003} or be structured radially as a two velocity flow \citep{Ghisellini2005}.
Recently, by considering that the apparent motion of individual components result from some pattern motion, 
such as a shock wave or a plasma instability propagating in the jet,
\cite{Plavin2019} argued that the true flow velocity in the jet may be higher than the one estimated from the observed apparent motions.
Interestingly, using 13 years of observations with the \emph{Swift-XRT}, \cite{Hervet2019} confirmed that the variability pattern in Mrk 421 is consistent with a perturbation passing through a recollimation shock suggested by \cite{Marscher1985}.
They found that the deduced Lorentz and Doppler factors of the flow are relatively high, and are within the range $\Gamma\in[43-66]$ and $\delta\ge31$, respectively.
Thus, it seems that the shock acceleration may be in favor.

Alternatively, the high $\delta_D-$values can be easily produced in the scenario that an emitter moves relativistically inside a relativistic larger-scale jet traveling towards the observer\citep[][and references therein]{Giannios2009,Aharonian2017}.
However, the most feasible energy source for this motion is magnetic field reconnection in a highly magnetized jet, which may be disfavored by our results (see discussion below).

From the modelling results, we also inferred the kinetic power $L_{\rm kin}$, magnetization $\sigma_{\rm m}$ and radiative efficiency $\eta_{\rm r}$ of the
jets with a pure $e^\pm$-pairs ($\eta_{\rm p}=0$) and electron-proton ($\eta_{\rm p}=1$) composition in the framework of the three models.
The values and $1\sigma$ errors of the derived parameters for both $\eta_{\rm p}=0$ and $\eta_{\rm p}=1$ are reported in Table \ref{tab:source3}.
The complete information on the constraining parameters are presented in Figure \ref{2dpdfA} in the Appendix \ref{two-dimensional contours}.

In Figure \ref{kin}, the dependence of $L_{\rm kin}$, $\sigma_{\rm m}$ and $\eta_{\rm r}$ on the minimum variability timescale $t_{\rm v,min}$ are shown,
where the timescale $t_{\rm v,min}$ is related to the size of the source through the causality relation and is calculated by $t_{\rm v,min}=R'(1+z)/c\delta_{\rm D}$.
It can be found that the derived variability timescales can be ranged from $\sim2$ hours to $\sim1.5$ months.
The family of solutions obtained with our approach also are shown in Figure \ref{kin}.
In the calculation, Model A2 is adopted to be the benchmark model.
It can be seen that the values and variations of our interesting physical quantities over a wide ranges of $t_{\rm v,min}$ can be well reproduced by our approach,
and their values derived with $\eta_{\rm p}=0$ and $\eta_{\rm p}=1$ are compatible with each other within the 1$\sigma$ errors.
Depending on the variability timescale and plasma composition, $L_{\rm kin}$ varies only by a factor of $\sim2-3$,
while both $\sigma_{\rm m}$ and $\eta_{\rm r}$ vary by more than an order of magnitude.
The impact of the plasma composition decreases with the increasing $t_{\rm v,min}$ from hours to months.
These results indicate that it is difficult to exactly determine the value of the parameter $\eta_{\rm p}$, which can span the range of $0-1$.

Finally, we find that jet kinetic power $L_{\rm kin}$ of Mrk 421 narrowly range from $\sim1.6\times10^{44}$ to $\sim4.0\times10^{44}$ erg/s ,
while $\sigma_{\rm m}$ and $\eta_{\rm r}$ can be ranged from 0.01 to 0.3 and from 0.3\% to 10\%, respectively.
Compared to Mrk 421, it can be found that $L_{\rm kin}$ in Mrk 501 are significantly larger,
while $\sigma_{\rm m}$ and $\eta_{\rm r}$ are significantly lower by more than one order of magnitude,
i.e. for Mrk 501, $L_{\rm kin}$ narrowly range from $\sim4.0\times10^{44}$ to $\sim10^{45}$ erg/s,
while $\sigma_{\rm m}$ and $\eta_{\rm r}$ range from 0.0003 to 0.01 and from 0.03\% to 1\%.

\begin{figure*}
  \centering
  \includegraphics[width=0.45\textwidth]{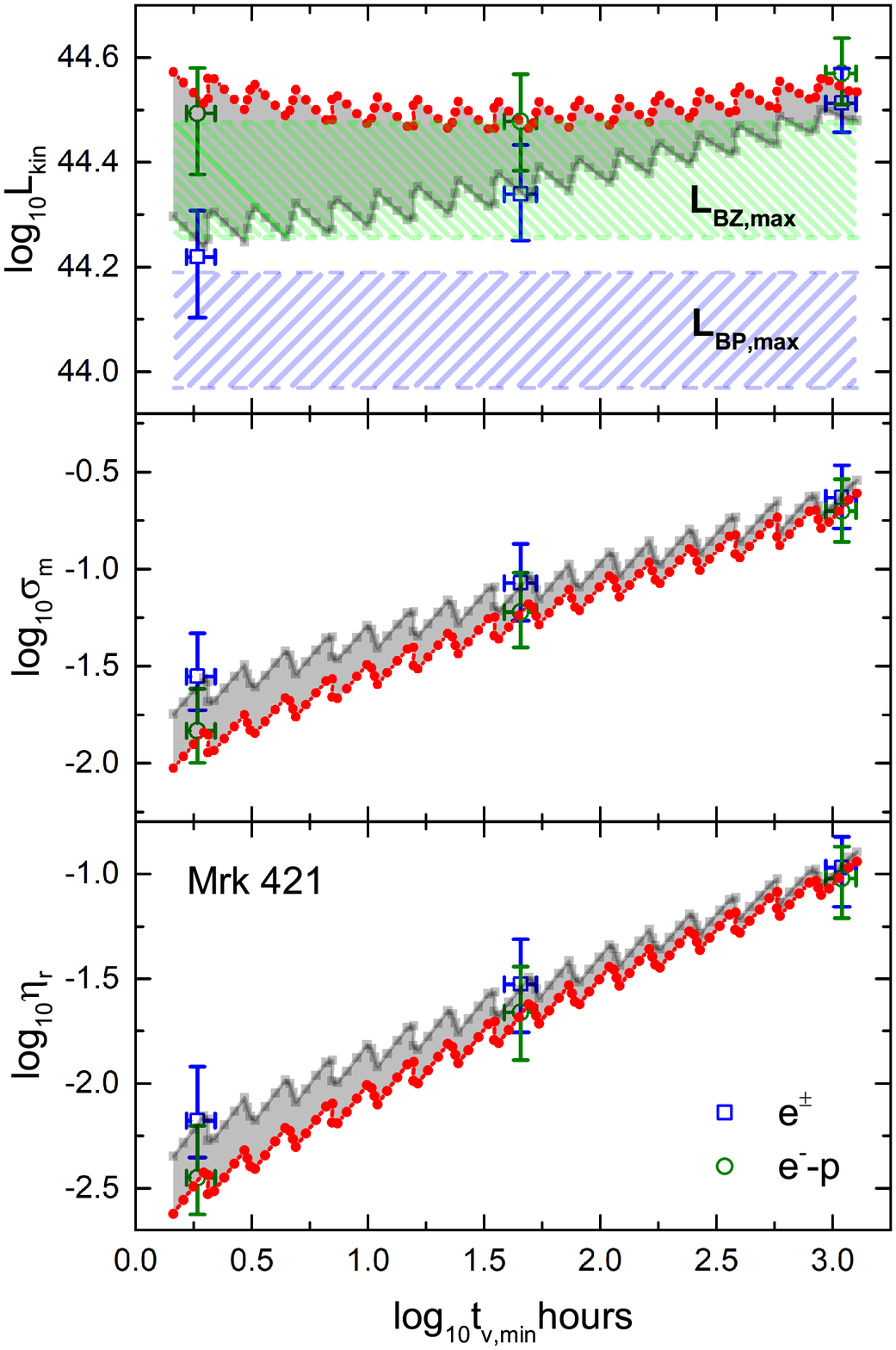}\includegraphics[width=0.45\textwidth]{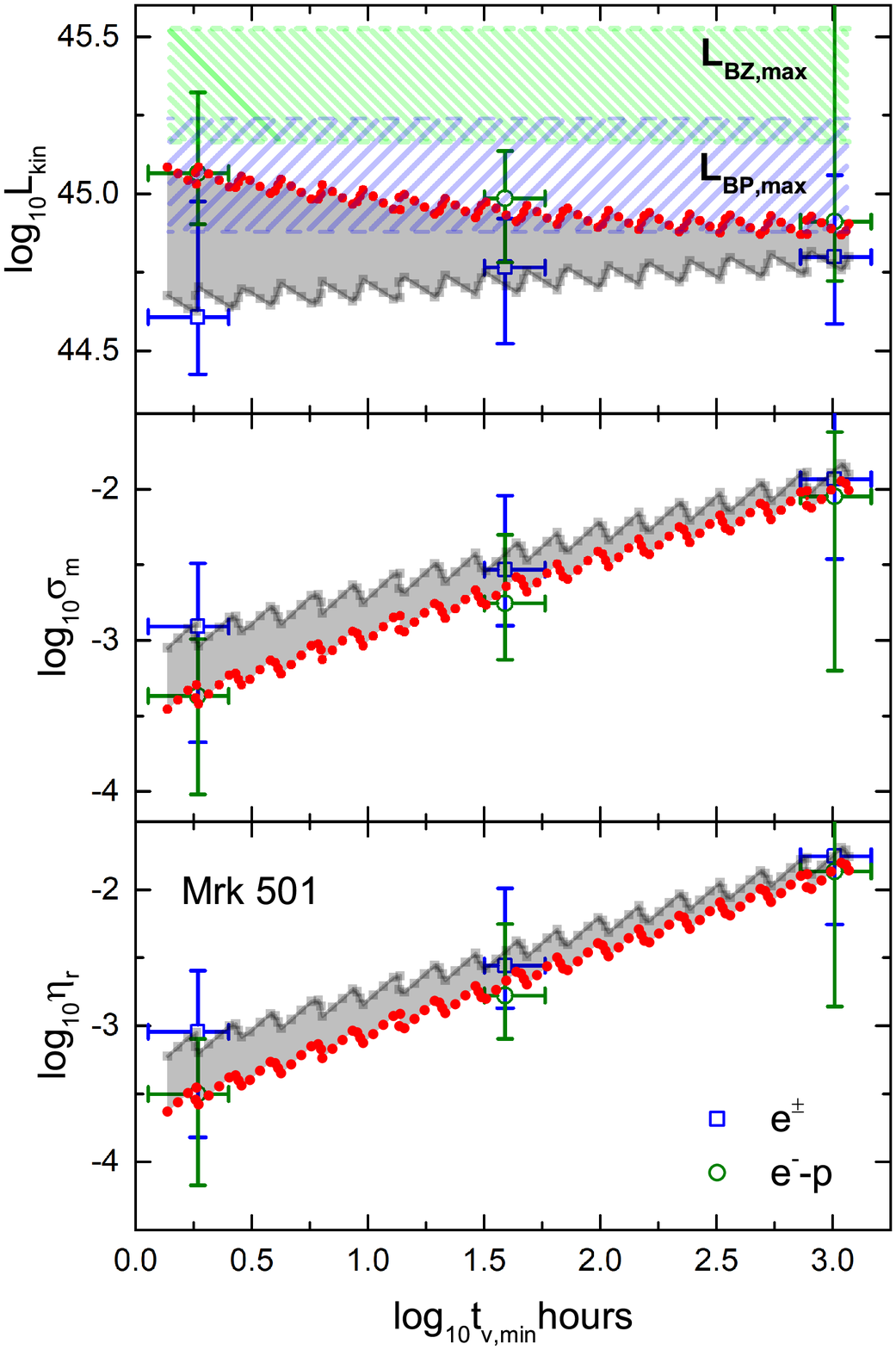}
  \caption{The jet kinetic power $L_{\rm kin}$ (Top), magnetization $\sigma_{\rm m}$ (middle) and radiative efficiency $\eta_{\rm r}$ (bottom) as a function of the minimum variability timescales $t_{\rm v,min}$.
  The opened blue squares and olive circles are the results derived by fitting the SED under the assumption of pure pairs and electron-proton plasma, respectively.
  The filled gray squares and red circles denote the family of solutions predicted by our method for pure pairs and electron-proton plasma, respectively.
 }\label{kin}
\end{figure*}

\begin{table*}
\def\arraystretch{1.5}
\caption{The values of the derived parameters for Mrk 421 and Mrk 501.}
\label{tab:source3}
\begin{tabular}{lcccccccc}
\hline
Model & $\log_{10}L_{\rm kin,e^\pm}$	  & $\log_{10}L_{\rm kin,e^--p}$	& $\log_{10}\sigma_{\rm e^\pm}$ & $\log_{10}\sigma_{\rm e^--p}$	&  $\log_{10}\eta_{\rm r,e^\pm}$  & $\log_{10}\eta_{\rm r,e^--p}$  & $\log_{10}t_{\rm min}(hr)$  \\
     \hline
A1& $44.22_{-0.12}^{+0.09}$ & $44.49_{-0.12}^{+0.09}$& $-1.55_{-0.17}^{+0.22}$ &$-1.83_{-0.16}^{+0.22}$ &$-2.18_{-0.18}^{+0.26}$& $-2.45_{-0.18}^{+0.25}$   &$0.27_{-0.05}^{+0.08}$\\
A2& $44.34_{-0.09}^{+0.09}$ & $44.48_{-0.09}^{+0.09}$& $-1.07_{-0.19}^{+0.20}$ &$-1.22_{-0.18}^{+0.20}$ &$-1.53_{-0.23}^{+0.22}$& $-1.66_{-0.23}^{+0.22}$  &$1.66_{-0.07}^{+0.07}$\\
A3& $44.51_{-0.06}^{+0.07}$ & $44.57_{-0.06}^{+0.07}$& $-0.63_{-0.16}^{+0.17}$ &$-0.70_{-0.16}^{+0.16}$ &$-0.97_{-0.18}^{+0.15}$& $-1.02_{-0.19}^{+0.15}$  &$3.04_{-0.07}^{+0.06}$\\
\hline
A1& $44.61_{-0.18}^{+0.37}$ & $45.06_{-0.16}^{+0.26}$& $-2.91_{-0.76}^{+0.42}$ &$-3.37_{-0.65}^{+0.38}$&$-3.04_{-0.78}^{+0.45}$& $-3.50_{-0.67}^{+0.41}$&$0.27_{-0.22}^{+0.13}$ \\
A2& $44.76_{-0.24}^{+0.15}$ & $44.98_{-0.20}^{+0.15}$& $-2.53_{-0.37}^{+0.49}$ &$-2.75_{-0.37}^{+0.45}$&$-2.78_{-0.32}^{+0.52}$& $-2.78_{-0.32}^{+0.52}$&$1.59_{-0.09}^{+0.17}$ \\
A3& $44.80_{-0.21}^{+0.26}$ & $44.91_{-0.19}^{+1.11}$& $-1.93_{-0.53}^{+0.45}$ &$-2.05_{-1.15}^{+0.43}$&$-1.75_{-0.50}^{+0.49}$& $-1.86_{-0.99}^{+0.47}$&$3.01_{-0.15}^{+0.16}$ \\
\hline
\end{tabular}
\begin{flushleft}
Note: quantities derived with $\eta_{\rm p}=0$ are denoted by the subscript `$e^\pm$',
and ones derived with $\eta_{\rm p}=1$ are denoted by the subscript `$e^--p$'. \\
\end{flushleft}
\end{table*}

\subsection{Jet formation mechanisms}

It is generally believed that a relativistic jet of BL Lacs can be launched through either the Blandford-Payne \citep[BP;][]{Blandford1982}
and/or Blandford-Znajek \citep[BZ;][]{Blandford1977} mechanisms.
In the former scenario, the jet energy is governed by the gravitational energy released from the matter that accretes towards the BH,
while the rotational energy of a rapidly rotating BH is essential to drive a jet in the latter scenario.

For a central BH with a mass of $M_{\rm BH}$, the accretion power can be expressed as
\begin{equation}
L_{\rm acc}=\dot{m}\dot{M}_{\rm Edd}c^2=1.26\times10^{46}\dot{m}M_8, ~\rm ergs/s
\end{equation}
where $\dot{m}$ is the dimensionless mass accretion rate, $M_8$ is the BH mass in units of $10^8M_\odot$($M_\odot$ is the solar mass),
and $\dot{M}_{\rm Edd}=L_{\rm Edd}/c^2$ is the Eddington accretion rate.
Note that the jet power predicted by the BP mechanism can not exceed $L_{\rm acc}$, i.e. $L_{\rm BP}\lesssim L_{\rm BP,max}=L_{\rm acc}$ \citep{Blandford1982,Jolley2009,Ghisellini2010}.

In the magnetically arrested/choked accretion flows \citep{Blandford1977,Tchekhovskoy2010,Tchekhovskoy2011},
the rate of the energy extraction from a rotating BH via the BZ process can be given by
\begin{equation}\label{Lbz}
L_{\rm BZ}\simeq1.26\times10^{47}\left(\phi_{\rm BH}/50\right)^2x_a^2f(x_a)\dot{m}M_8, ~\rm ergs/s
\end{equation}
where $\phi_{\rm BH}$ is the dimensionless magnetic flux threading a BH,
$x_a\equiv a/[2(1+\sqrt{1-a^2})]$, $f(x_a)\simeq1+1.4x_a^2-9.2x_a^4$, with $a$ denoting the dimensionless angular momentum parameter (also called ``spin'') \citep{Sikora2013}.
When the typical values of $\phi_{\rm BH}=50$ is adopted \citep{McKinney2012},
the predicted jet power is $L_{\rm BZ,max}\simeq2.44\times10^{46}\dot{m}M_8$ for maximal BH spins, i.e. $a=1$.
This value is about 1.9 times of the $L_{\rm acc}$.

Both $L_{\rm BP,max}$ and $L_{\rm BZ,max}$ are mainly determined by the fundamental parameters: $M_{\rm BH}$ and $\dot{m}$.
In the following, we adopt the BH mass estimated from the direct measurement of stellar velocity dispersion,
which is widely used to estimate the BH mass in BL Lacs \citep[see, e.g.][]{Woo2002}.
Using the method, the mass of the BH hosted in Mrk 421 and Mrk 501
are $\log M_{\rm BH}/M_\odot=8.28\pm0.11$ and $\log M_{\rm BH}/M_\odot=9.21\pm0.13$, respectively \citep{Woo2002,Barth2003}.

Note that the mass accretion rate in BL Lacs is known to be quite low and the accretion flow is likely in the the radiative-inefficient regime \citep{Wang2002,Xu2009,Ghisellini2010}.
For HBLs, the typical value of $\dot{m}$ is about $5\times10^{-3}$ \citep{GT2008,Sbarrato2012,Sbarrato2014}.
Thus, we obtained that $L_{\rm BP,max}$ for Mrk 421 and Mrk 501 range from $9.3\times10^{43}$ to $15.5\times10^{43}$ erg/s and range from $7.6\times10^{44}$ to $17.4\times10^{44}$ erg/s, respectively.
On the other hand,
$L_{\rm BZ,max}$ is in range of $(1.8-3.0)\times10^{44}$ erg/s and in range of $(1.5-3.4)\times10^{45}$ erg/s for Mrk 421 and Mrk 501, respectively.
For comparison, both $L_{\rm BP,max}$ and $L_{\rm BZ,max}$ are shown in the $t_{\rm v,min}-L_{\rm kin}$ plots for Mrk 421 and Mrk 501, respectively (see Figure \ref{kin}).

Since the total jet power estimated from the SED fitting is dominated by the kinetic power $L_{\rm kin}$, the value of the the total jet power can be approximately equal to $L_{\rm kin}$.
It can be seen from Figure \ref{kin} that for Mrk 421 $L_{\rm kin}$ is roughly consistent with $L_{\rm BZ,max}$, but seems to be difficult to reconcile with $L_{\rm BP,max}$.
This implies that the accretion power may be not sufficient to launch the jets in Mrk 421,
and the BZ mechanism may be in favour.

For Mrk 501, it can be found that $L_{\rm BZ,max}$ is systematically higher than $L_{\rm kin}$, and $L_{\rm kin}$ is comparable with $L_{\rm BP,max}$,
implying that the jet in Mrk 501 can be driven by either BZ or BP mechanism.
Under the BZ dominant mechanism, there may exist a relatively slowly spinning BH.
For the BP mechanism, a fraction $\sim0.7$ of $L_{\rm acc}$ may be required to produce the jet power.

The scenario of a pure accretion-driven jet may be not in favour due to that for a sample of BL Lacs
a weak anti-correlation between the jet power and the BH mass was observed in \cite{Zhang2012}.
They suggested that the spin energy of the central BH should have a significant role to
play in the production of the jets, and the weak anti-correlation implies a decrease in the BH spin with an increase in the BH mass.

It should be noted that the mass accretion rate $\dot{m}$ is largely uncertain for BL Lacs.
On the one hand, there are no direct signatures of the accretion, since the observed continuum emission from the jets is strongly beamed to us.
On the other hand, perturbations in the accretion rate are expected to occur in an underlying accretion disk \citep[e.g.,][]{Lyubarskii1997,Cowperthwaite2014}. 
However, there is some evidence to suggest that ADAFs may be in most Fanaroff-Riley I type radio galaxies(FR Is) \citep[see][and references therein]{Wu2008},
which are believed to be BL Lacs with the relativistic jet misaligned to our line of sight.
Based on the assumption that most of the X-ray emission is from the ADAFs,
 \cite{Wu2011} found that $\dot{m}$ can be limited in the range of $\sim10^{-4}$ to $10^{-2}$ for a sample of the FR Is.
With a sample composed of blazars and radio galaxies, \cite{Sbarrato2014} identified the transition of accretion flow from a standard Shakura-Sunyaev disk with a radiatively inefficient disk (e.g., ADAF), and found that such a transition occurs at $L_{\rm BLR}/L_{\rm Edd}\sim5\times10^{-4}-10^{-3}$
with $L_{\rm BLR}$ the luminosity of the broad-line region,
i.e., $\dot{m}\sim5\times10^{-1}-0.1$ when a radiative efficiency of $\eta\sim 0.1$ and an average covering factor of $f_{\rm cov}\sim0.1$ are assumed.
Particularly, one can find $L_{\rm BLR}/L_{\rm Edd}\simeq10^{-5}$ for Mrk 421 and Mrk 501 \citep[see][]{Ghisellini2011,Sbarrato2012}.
In other words, the mass accretion rates for these two HBLs can be estimated to be $\dot{m}\simeq10^{-3}$.

\begin{figure}
  \centering
  \includegraphics[width=0.45\textwidth]{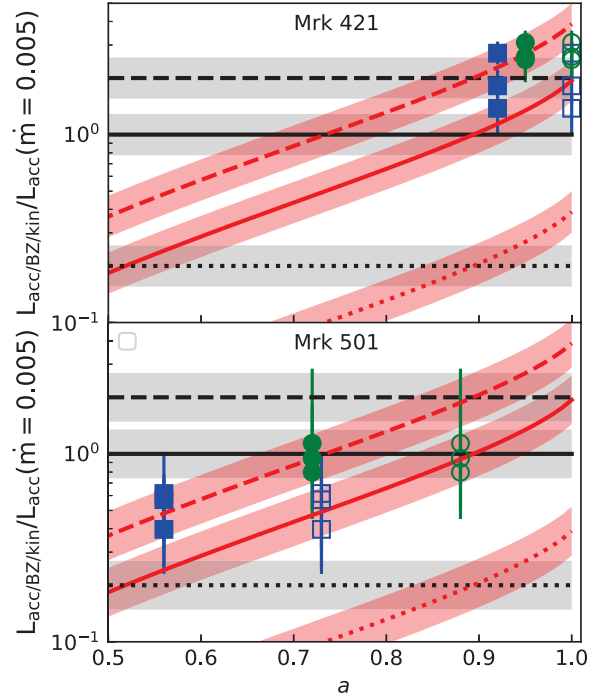}
  \caption{
$L_{\rm BZ}$ as a function of the BH spin $a$.
The dashed, solid and dotted red lines correspond to $\dot{m}=10^{-2}, 5\times10^{-3}$ and $10^{-3}$, respectively.
For the three given values of $\dot{m}$, $L_{\rm acc}$ are denoted by the dashed, solid and dotted black lines, respectively.
The colored bands around the lines represent the uncertainties from the BH mass $M_{\rm BH}$.
$L_{kin,e^\pm}$ and $L_{kin,e^--p}$ estimated from SED fittings are denoted by the blue squares and olive circles, respectively.
The opened and filled symbols represent $L_{kin,e^\pm}$($L_{kin,e^--p}$) at the minimal values of BH spin $a$ allowed by the BZ mechanism for $\dot{m}=5\times10^{-3}$ and  $\dot{m}=10^{-2}$, respectively.
Note that $L_{\rm acc}$, $L_{\rm BZ}$, and $L_{\rm kin}$ are normalized by $L_{\rm acc}(\dot{m}=5\times10^{-3})$.
The data points are the same as that in the top panels of Figure \ref{kin}.
 }\label{jetpower}
\end{figure}

Motivated by the above results, $\dot{m}=10^{-3}$, $\dot{m}=5\times10^{-3}$ and $\dot{m}=10^{-2}$ are taken to show the impact of changes in the accretion rate.
In Figure \ref{jetpower}, we show the jet power predicted by the BZ mechanism $L_{\rm BZ}$ as a function of the BH spin $a$,
and the accretion power $L_{\rm acc}$ and the kinetic power $L_{\rm kin}$ estimated from SED fitting are also shown for comparison.
It can be seen that the minimum spin parameter allowed by the BZ mechanism is $\gtrsim0.73-0.88$ for Mrk 501 in the case of $\dot{m}=5\times10^{-3}$,
while it may be required to reach maximum for Mrk 421.
In the case of $\dot{m}=10^{-2}$, the BH spin may be $\gtrsim0.56-0.72$ for Mrk 501 and $\gtrsim0.92-0.95$ for Mrk 421.
For a larger $\dot{m}$, we can obtain a smaller spin parameter under the BZ mechanism. However, the BP mechanism may be enough.
In the case of $\dot{m}=10^{-3}$, it seems that the BZ mechanism is insufficient to interpret $L_{\rm kin}$ estimated from SED fitting in particular for Mrk 421.
As long as the accretion rates are comparable in the two HBLs, we, therefore, expect that in Mrk 501 the BH spin is allowed to be smaller than that in Mrk 421
for any given values of $\dot{m}\gtrsim5\times10^{-3}$. 
Meanwhile, we stress that in Mrk 501 the minimal allowed value of $\dot{m}$ may be required to be lower than that in Mrk 421, if the BHs spin in the two HBLs could be comparable.

\subsection{Connection between the blazar zone and extended jet}

The kinetic power of jet can be estimated directly from the observations of the large-scale structures of radio galaxies \citep{Dunn2006,Birzan2008,Kino2012},
and some empirical relations between kinetic power and extended radio emission are built \citep{Willott1999,Cavagnolo2010,Meyer2011,Godfrey2013,Ineson2017}.
The kinetic power $L_{\rm kin,rl}$ can be inferred through the empirical relation, and is thus given by
\begin{equation}\label{kinjet}
 L_{\rm kin,rl} = g\times1.5\times10^{44}\left(\frac{L_{151}}{10^{32} \rm erg/s/Hz/sr}\right)^{0.67},
\end{equation}
where $L_{151}$ is the 151 MHz radio luminosity, and the normalization factor $g$ reflects the uncertain physics in lobes, including the
composition, magnetic field strength, electron spectrum, the bulk velocity of the hotspots plasma \citep{Godfrey2013}, can varies in the range of $\sim[1-8]$.

By applying the relation to several FR II radio galaxies with independent jet power measurements, a value of $g\simeq2$ was proposed by \citet{Godfrey2013}.
However, the low value may be not appropriate for FR I/BL Lacs because of the vastly different energy budgets between FR II/FSRQs and FR I/BL Lacs.
Thus, a high value of $g=8$ is also taken into account in our analysis.
Since the dissipation region of the jets in the two HBLs significantly depart from equipartition between the magnetic field and relativistic electrons.
Combined with the result shown in Figure 1 in \cite{Godfrey2013}, we assume that $g=2$ and $g=8$ may be responsible for the pure $e^\pm$ pairs and $e^--p$ compositions, respectively.

On the other hand, $L_{151}$ can be calculated through the relation $L_{151}=d_L^2F_{151}$,
where the 151 MHz radio flux $F_{151}=1.68\pm0.05$ Jy for Mrk 421 and $F_{151}=1.98\pm0.05$ Jy for Mrk 501 are taken from
NED\footnote{\url{http://ned.ipac.caltech.edu/forms/byname.html}}.
Omitting the small observation errors,
we find that for Mrk 421 and Mrk 501 $L_{\rm kin,rl}$ are respectively $2.8\times10^{43}$ and $3.6\times10^{43}$ erg/s when $g=2$,
and are respectively $11.4\times10^{43}$ erg/s and $14.5\times10^{43}$ erg/s when $g=8$.

Combing the benefit of the MCMC method with our semi-analytical method presented in Appendix \ref{para_space_study}, we could confidently assume that
the real jet power should be in the region bounded by the powers of jet with the pure $e^\pm$ pair and $e^--p$ plasma (see figure \ref{kin}).
We find that, for Mrk 421, when the factor $g$ is adopted to be 2, $L_{\rm kin,rl}$ is a factor of $\sim6$ lower than $L_{\rm kin,e^\pm}$,
while $L_{\rm kin,rl}$ is a factor of $\sim3$ lower than $L_{\rm kin,e^--p}$ when $g$ is assumed to be 8.
In Mrk 501, $L_{\rm kin,rl}$ derived with $g=2$ and $g=8$ are a factor of $\sim11$ and $\sim8$ lower than $L_{\rm kin,e^\pm}$ and $L_{\rm kin,e^--p}$, respectively.
The discrepancy between the estimated jet power from SED fitting and low-frequency radio emission 
implies that a faction of the jet energy may be lost after leaving the dissipation region of the blazar.

These results may be supported by the numerical modelling of the evolution of radio galaxy lobes presented in \cite{Hardcastle2013}.
The authors have shown that the work done by the expanding radio lobes on the external environment is roughly equal to the energy stored in the lobes once the lobes are well established.
This can lead to the result that the time-averaged jet power from the extended radio luminosity
 is likely to be underestimated by a factor of a few (up to an order of magnitude in extreme cases).
Thus, we conclude that in Mrk 501 the jet energy losses from expanding radio structure may be more significant than that in Mrk 421.

\section{Conclusions}\label{sec:summa}

Using the standard one-zone SSC model and MCMC technique, we systematically analyze the most detailed quasi-simultaneous MWL SEDs of HBLs Mrk 421 and Mrk 501.
With the assumption of a pure $e^\pm$ pair and $e^--p$ plasma, we inferred $L_{\rm kin}$, $\sigma_m$ and $\eta_r$ through the SED fitting.
Our results indicate that in Mrk 421 $L_{\rm kin}$ is lower, but $\sigma_{\rm m}$ and $\eta_{\rm r}$ are larger than that in Mrk 501.
The plasma composition in the jets of the two sources are difficult to constrain in the framework of our model, since the jet properties derived with
a pure $e^\pm$ pair and $e^--p$ plasma are consistent with each other within $1\sigma$ CLs.

Compared to Mrk 501, the higher $\sigma_{\rm m}$ jet of Mrk 421 implies that the energy could be more efficiently transported to a large scale  \citep[][]{Chen2018}.
On the other hand, the low $\eta_{\rm r}$ in these two sources are consistent with the prediction of the internal shock model \citep{Spada2001,Zhang2011}\footnote{In the models, the radiative efficiency is generally less than 15\%.},
and imply that only a small fraction of the jet energy is dissipated before being transported to a large scale.

By comparing the jet power estimated from the SED fitting and the accretion power of the central BH,
we find that the jets in both Mrk 421 and Mrk 501 may be powered by rapidly rotating BHs.
Based on the assumption of BZ mechanism, the spin of the BH in Mrk 501 is allowed to be lower, but it is also possible that they are equal.

The comparisons of the jet power estimated from SED fitting and from observation of the large-scale radio structure, showing that
the jet energy losses in the propagation between the blazar zone and the large-scale radio structure
may be important, and the energy losses in Mrk 421 is less significant than that in Mrk 501.

\section*{Acknowledgements}
We thank the anonymous referee for valuable suggestions.
We acknowledge financial support from the National Key R\&D Program of China (grant No. 2018YFA0404204) and
the National Natural Science Foundation of China (grant No. NSFC-11803027, NSFC-U1831124 and NSFC-12065011).
BZD acknowledges funding supports from the Science Foundation of Yunnan Province (grant No. 2018FA004).
WH acknowledges funding supports from the Key Laboratory of Astroparticle Physics of Yunnan Province (No. 2016DG006).

\section*{DATA AVAILABILITY}
The data underlying this article will be shared on reasonable request to the corresponding author.







\appendix

\section{$\delta$-function approximation for the synchrotron radiation}

\label{appenda}
Using $\delta$-function approximation for the synchrotron radiation gives \citep[e.g.,][]{Dermer1997,DS2002,DM2009}
\begin{equation} 
f^{\rm syn}_{\epsilon,\delta}=f_0V_{\rm b}'B'^2\delta_{\rm D}^4\gamma'^3n_{\rm e}'(\gamma').
\end{equation}
where the factor $f_0=c\sigma_{\rm T}/48\pi^2d_{\rm L}^2$ with $\sigma_{\rm T}$ denoting the Thomson cross section,
and electron energy $\gamma'$ is related to the observed dimensionless photon energy $\epsilon$ by the relation
\begin{equation} 
\epsilon=(4/3)\epsilon_{\rm B}^\prime{\gamma'}^2\frac{\delta_{\rm D}}{1+z}.
\end{equation}

Then, the synchrotron peak frequency and peak flux can be expressed as
\begin{eqnarray}
  \nu_{\rm pk} &=& \nu_0 B'\delta_{\rm D}\gamma_{\rm pk}'^2,\label{nupk1} \\
  f_{\rm pk}^{\rm syn} &=& f_0 V_{\rm b'}B'^2\delta_{\rm D}^4\gamma_{\rm pk}'^3n_{\rm e}'(\gamma_{\rm pk}')\label{fsyn1}
\end{eqnarray}
where $\nu_0=4m_{\rm e}c^2/3hB_{\rm cr}(1+z)$ and $\gamma_{\rm pk}'$ denotes the electron Lorentz factor corresponding the synchrotron peak frequency.


\section{Study of the parameter space}
\label{para_space_study}
\begin{figure}
  \centering
  \includegraphics[width=8cm]{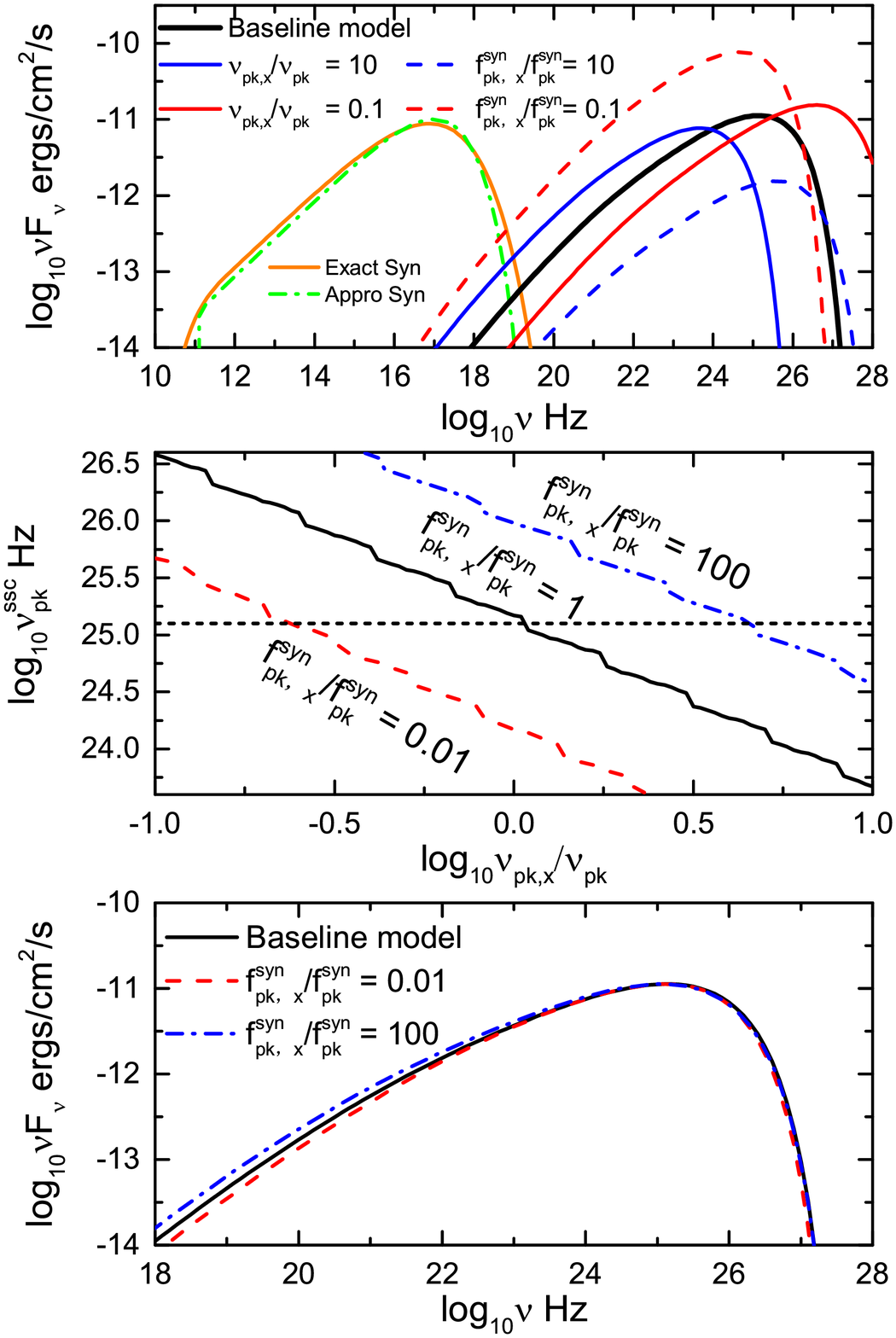}\\
  \caption{The upper panel displays the effects of changing $\nu_{\rm pk,x}$ and $f_{\rm pk,x}^{\rm syn}$.
The middle panel displays the relations between SSC peak frequency $\nu_{\rm pk}^{\rm ssc}$ and $\nu_{\rm pk,x}/\nu_{\rm pk}$.
 The horizontal dashed line denotes the SSC peak frequency of the benchmark model.
The bottom panel displays the SSC spectra reproduced with the solutions estimated in the middle panel.
}\label{base}
\end{figure}

In the section, we present a semi-analytical method that can lead to equally good models over essentially the entire range of values probed in the framework of the standard one-zone SSC model.

Based on the SSC mechanism, the intrinsic radius of the blazar zone $R'$ can be rewritten as
\begin{equation}\label{radius}
R'=[d_{\rm L}/(1+z)]\sqrt{\mathcal{F}_{\rm pk}^{\rm ssc}/f_{\rm pk,obs}^{\rm ssc}}.
\end{equation}
where $\mathcal{F}_{\rm pk}^{\rm ssc}$ denotes the peak flux of the function $\mathcal{F}_{\epsilon_\gamma}^{\rm ssc}$ (Eq. \ref{source_flux}) associated with the magnetic field $B'$ and doppler factor $\delta_{\rm D}$,
and $f_{\rm pk,obs}^{\rm ssc}$ denotes the peak flux of the observed high-energy bump.
Notice that the influence of SSA process on $\mathcal{F}_{\rm pk}^{\rm ssc}$ can be neglected.

Then, we introduce the quantities $\nu_{\rm pk,x}$ and $f_{\rm pk,x}^{\rm syn}$ as the free parameters instead of $B'$ and $\delta_{\rm D}$.
Here, $\nu_{\rm pk,x}$ and $f_{\rm pk,x}^{\rm syn}$ are defined as
\begin{eqnarray}
  \nu_{\rm pk,x} &=& \nu_0 B_x'\delta_{\rm D,x}\gamma_{\rm pk}'^2, \label{nupk2}\\
   f_{\rm pk,x}^{\rm syn} &=& f_0 V_{\rm b}'B_{\rm x'^2}\delta_{\rm D,x}^4\gamma_{\rm pk}'^3n_{\rm e'}(\gamma_{\rm pk}').\label{fsyn2}
\end{eqnarray}

By combining Eq. \ref{nupk1}, \ref{fsyn1}, \ref{nupk2} and \ref{fsyn2},
the values of $B'$ and $\delta_{\rm D}$ are updated by
\begin{eqnarray}
  B_{\rm x}' &=&\frac{(\nu_{\rm pk,x}/\nu_{\rm pk})^2}{\sqrt{f_{\rm pk,x}^{\rm syn}/f_{\rm pk}^{\rm syn}}}B',\label{input2}\\ 
  \delta_{\rm D,x} &=&  \frac{\sqrt{f_{\rm pk,x}^{\rm syn}/f_{\rm pk}^{\rm syn}}}{\nu_{\rm pk,x}/\nu_{\rm pk}}\delta_{\rm D}\label{input3}. 
\end{eqnarray}

Lastly, a alternative solution can be obtained by equating the peak frequency $\nu_{\rm pk}^{F}$ of the function $\mathcal{F}_{\epsilon_\gamma}^{\rm ssc}$ with $\nu_{\rm pk,obs}^{\rm ssc}$, i.e. $\nu_{\rm pk}^{F}=\nu_{\rm pk,obs}^{\rm ssc}$, where $\nu_{\rm pk}^{F}$ is obtained by taking the maximum of Eq. \ref{source_flux}.
Here, the equation is solved numerically by using the routine \texttt{RTBIS} from \cite{Press1992}.

For illustration, we perform a parameter study by varying $\nu_{\rm pk,x}$ and $f_{\rm pk,x}^{\rm ssc}$, when the remaining parameters characterizing the synchrotron spectrum have been frozen.
The numerical results are presented in Figure \ref{base}.
In this exercise, we consider a source located at redshift $z=0.034$ corresponding the luminosity distance $ d_{\rm L}=150.3~{\rm Mpc}$,
and assume that a best-fit solution for a given values of $R'=2\times10^{15}$ cm is given by:
$B'=0.1 ~{\rm G},~\delta_{\rm D}=20$, $\nu_{\rm pk}=10^{17} {\rm Hz},~f_{\rm pk}^{\rm syn}=10^{-11} {\rm erg/s/cm^2}, \eta_l=10^{-3}, \eta_{\rm u}=10^2,~\alpha=2,~\beta=1$.
It is referred as the benchmark model, which gives $f_{\rm pk,obs}^{\rm ssc}\simeq10^{-11}~ {\rm ergs/cm^2/s}$, $\nu_{\rm pk,obs}^{\rm ssc}\simeq10^{25}~{\rm Hz}$.

From the upper panel of Figure. \ref{base},
it can be seen that the Compton peak flux $f_{\rm pk}^{\rm ssc}$ increases slowly with the decreasing $\nu_{\rm pk,x}$,
while the Compton peak frequency $\nu_{\rm pk}^{\rm ssc}$ increases rapidly.
Contrary to the effects of changing $\nu_{\rm pk,x}$, $f_{\rm pk}^{\rm ssc}$ increases rapidly with the decreasing $f_{\rm pk,x}^{\rm syn}$,
while $\nu_{\rm pk}^{\rm ssc}$ decreases slowly.

In the middle panel of Figure. \ref{base}, we display the dependence of $\nu_{\rm pk}^{F}$ on $\nu_{\rm pk,x}$ for three values of $f_{\rm pk,x}^{\rm syn}/f_{\rm pk}^{\rm syn}=10^{-2},~1$ and $10^2$.
Obviously, $\nu_{\rm pk,x}=\nu_{\rm pk}$, when $f_{\rm pk,x}^{\rm syn}=f_{\rm pk}^{\rm syn}$. It corresponds to the benchmark model.
For the two other values of $f_{\rm pk,x}^{\rm syn}$, two solutions for $\nu_{\rm pk,x}$ can be found
by numerically solving $\nu_{\rm pk}^{F}=\nu_{\rm pk,obs}^{\rm ssc}$.

The two alternative solutions found with our approach, together with the benchmark model, are shown in the bottom panel of Figure. \ref{base}.
The results show that the SSC spectra around the peak frequencies are almost indistinguishable.
Compared to the benchmark model, the two alternative solutions are accurate to better than 5\% for the spectrum well below the peak,
which is smaller than the relative systematic uncertainty of 10\% for the $\gamma$-ray data \citep[e.g.,][]{Ackermann2012}.
Note that the difference is caused by variation of $\gamma_l'$, which plays a role on affecting the spectral slopes detected in the GeV band.
The family of solutions can be obtained with the parameter $f_{\rm pk,x}^{\rm syn}/f_{\rm pk}^{\rm syn}$ ranging from $10^{-2}$ to $10^2$.

\section{2D confidence contours}
\label{two-dimensional contours}

\begin{figure*}
  \centering
  \includegraphics[width=0.45\textwidth]{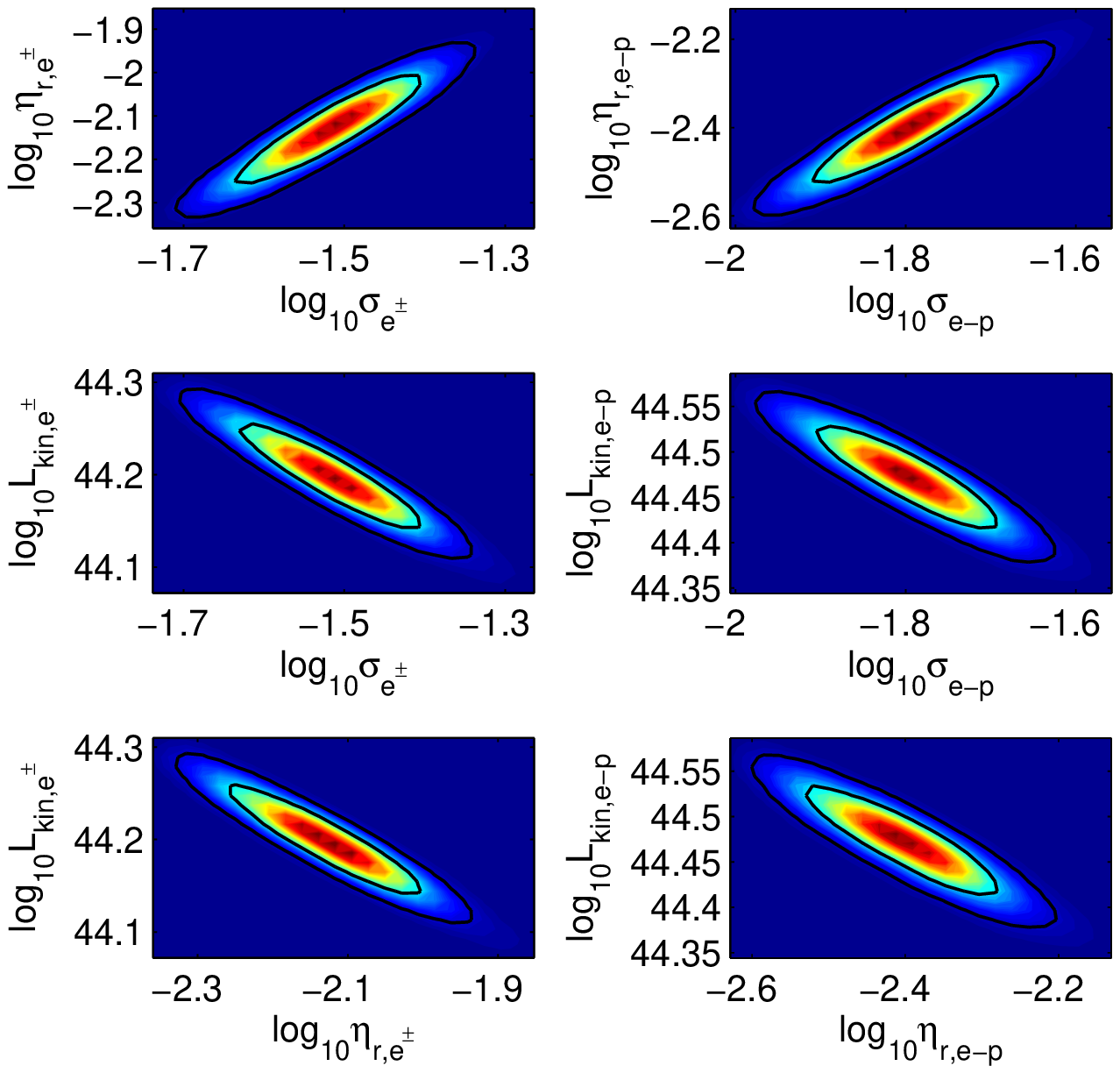} \includegraphics[width=0.45\textwidth]{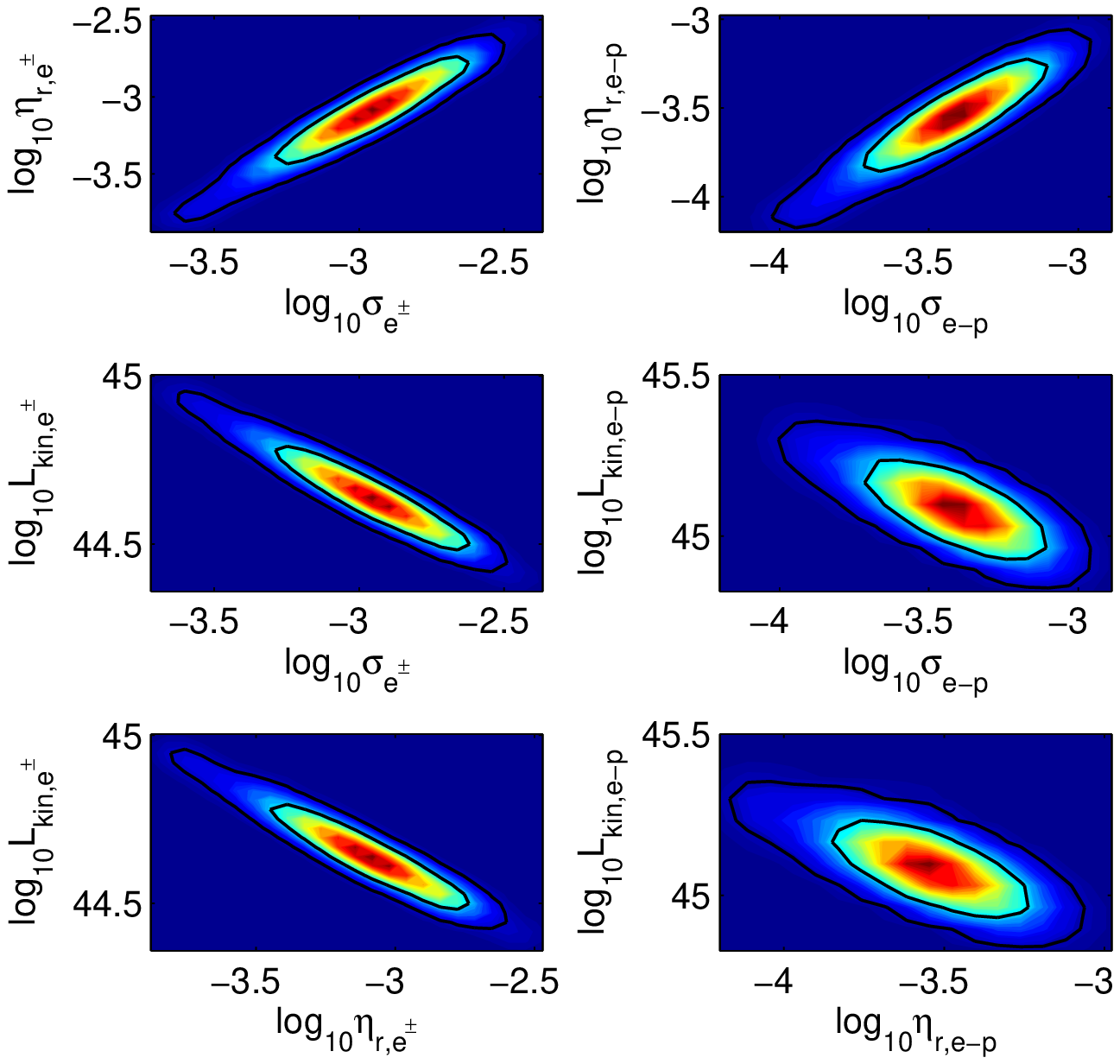}\\
  \includegraphics[width=0.45\textwidth]{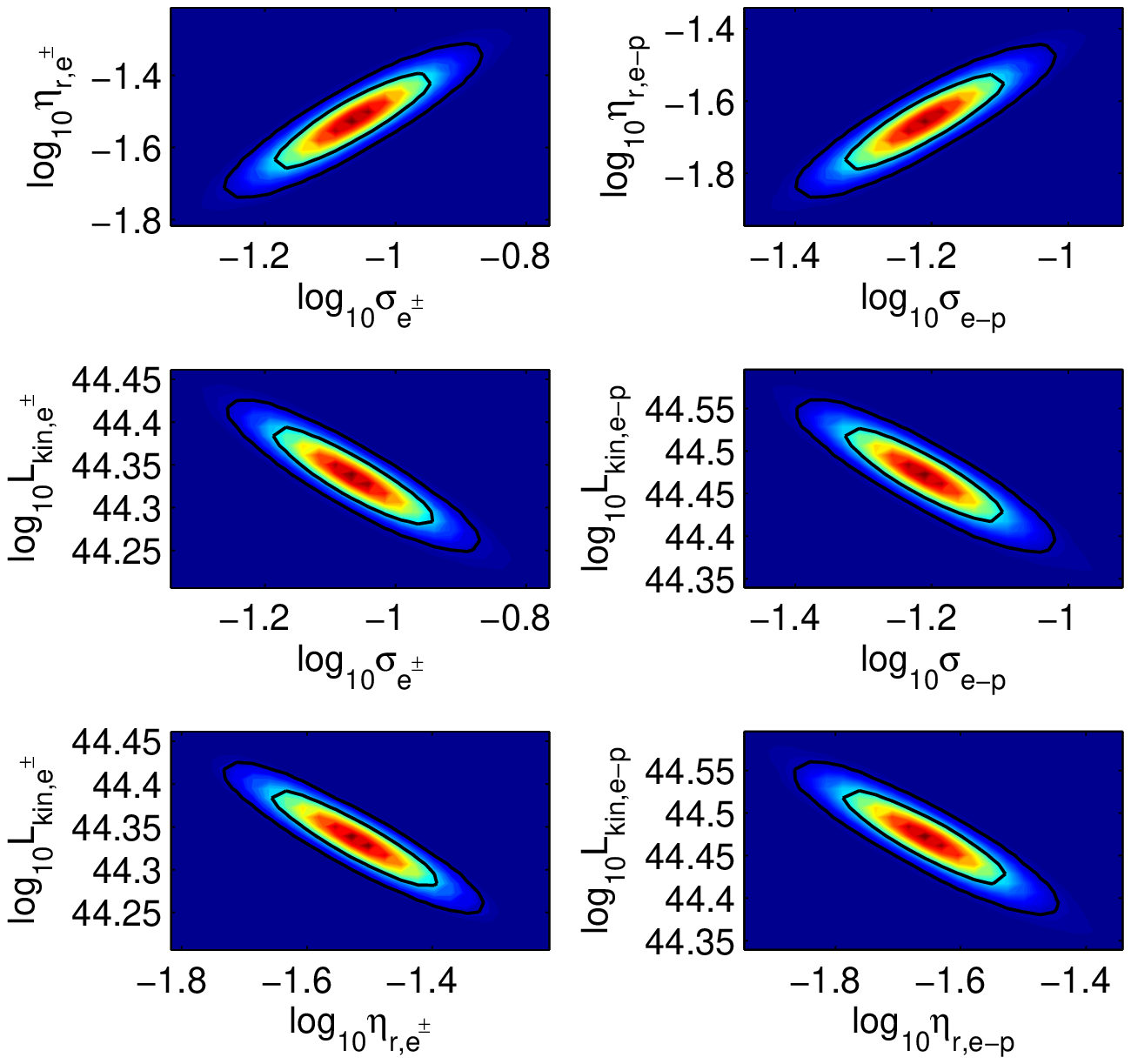} \includegraphics[width=0.45\textwidth]{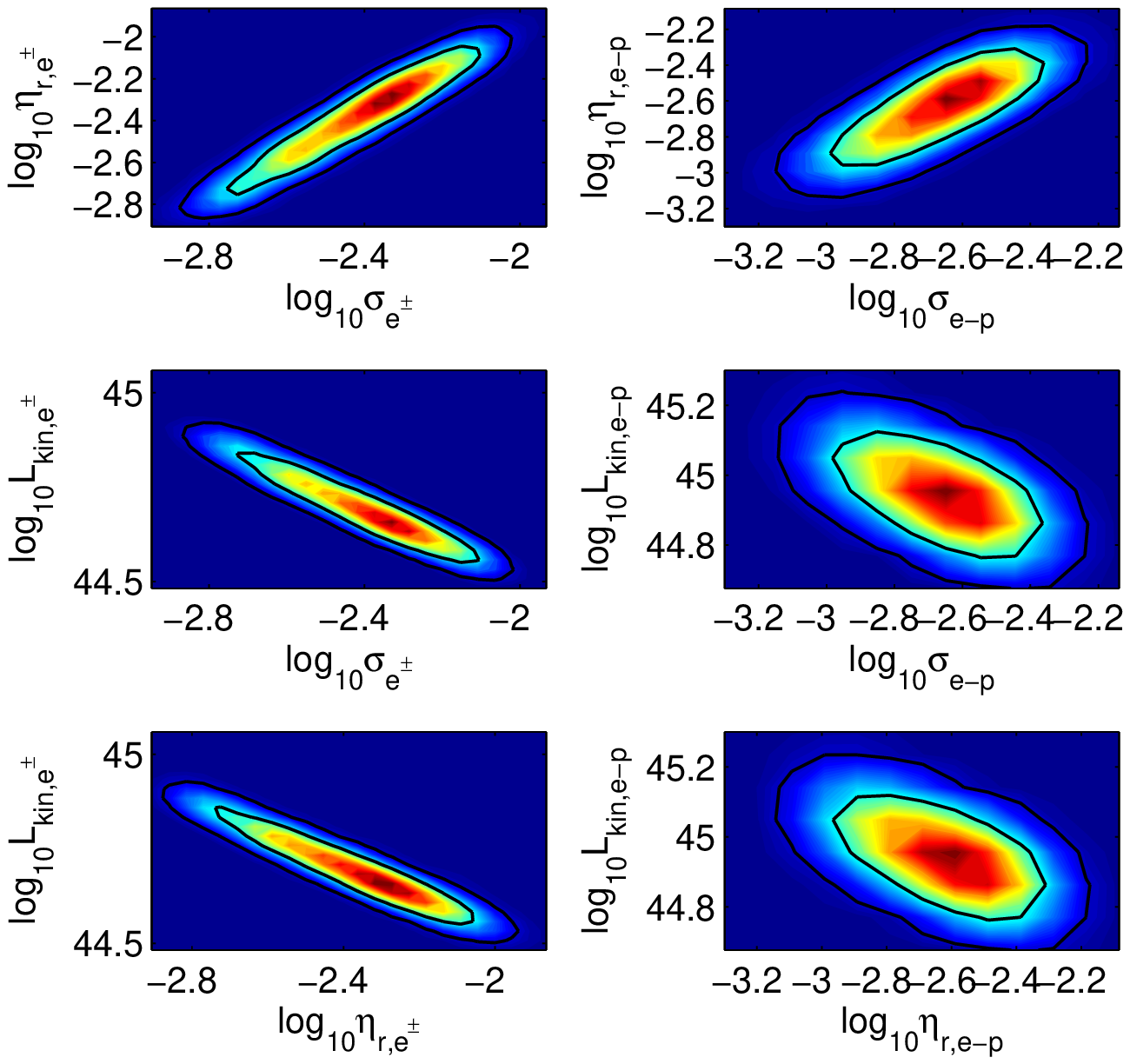}\\
  \includegraphics[width=0.45\textwidth]{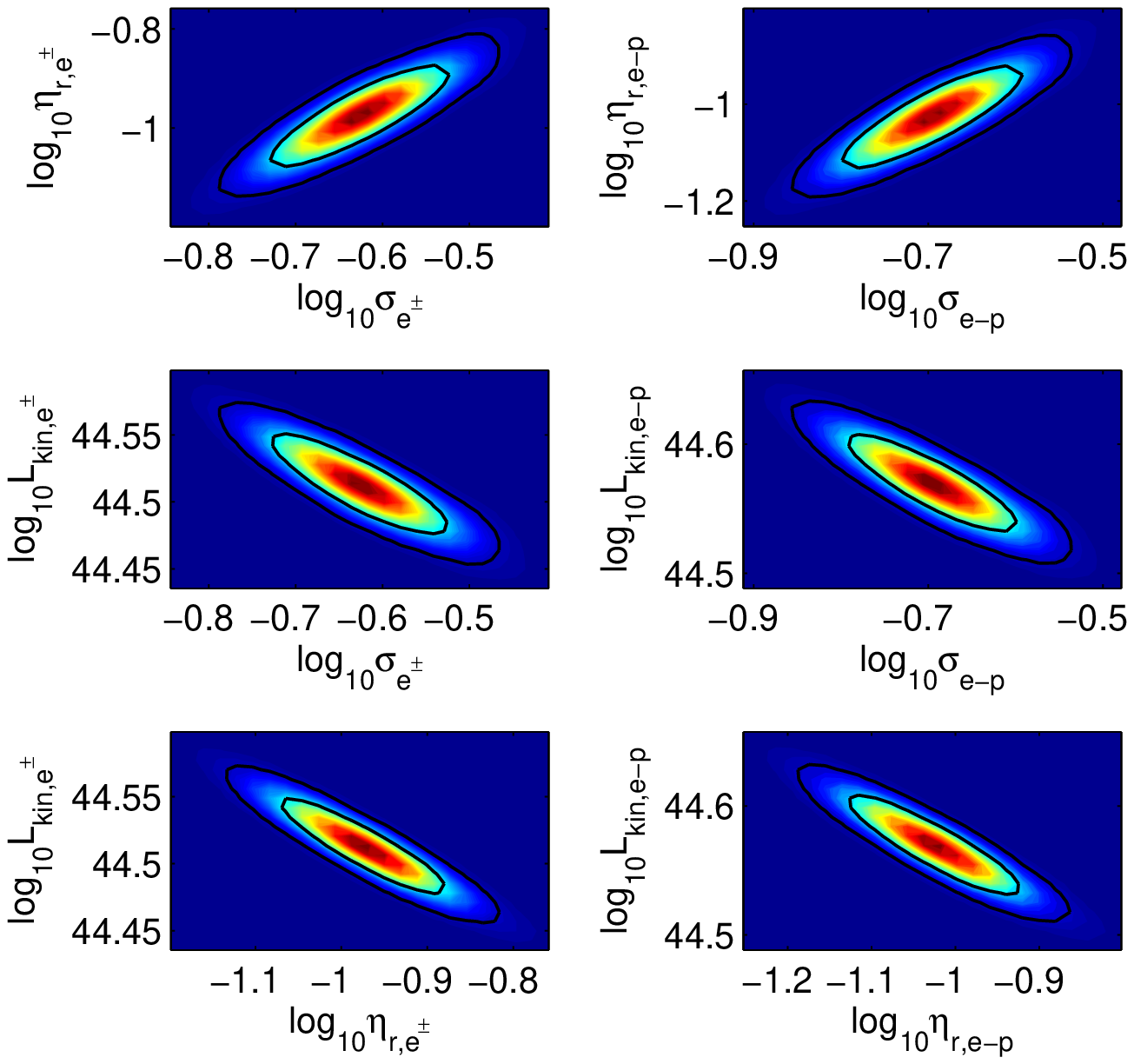} \includegraphics[width=0.45\textwidth]{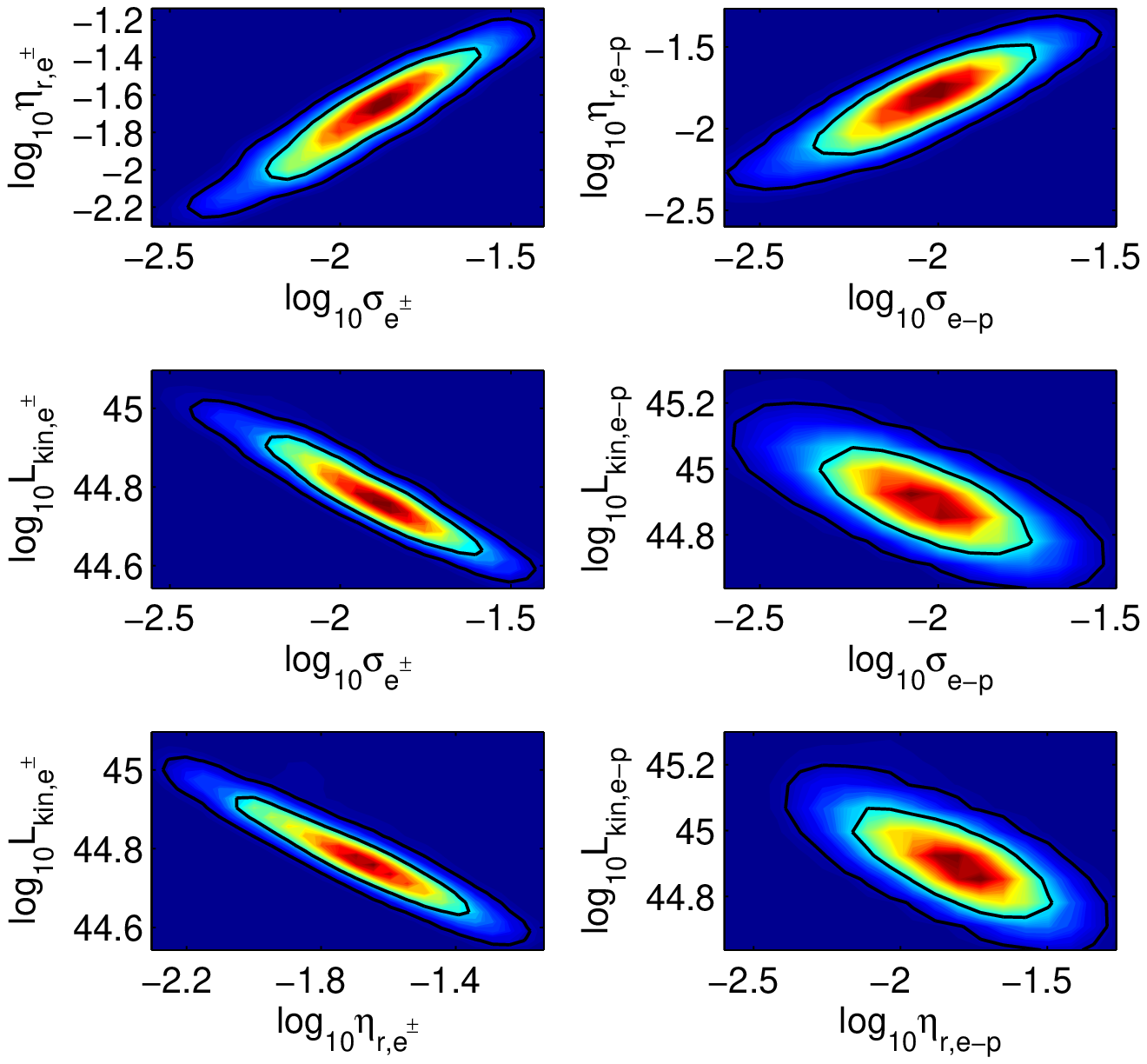}\\
  \caption{From top to bottom, 2D confidence contours of the derived parameters are for Model A1, A2 and A3, respectively. The left-hand panels are for Mrk 421, while the right-hand panels are for Mrk 501}\label{2dpdfA}
\end{figure*}

Figure \ref{2dpdfA} shows 2D confidence contours of the derived parameters,
which are obtained by using the MCMC code.

\bsp	
\label{lastpage}
\end{document}